\documentclass[graybox]{svmult}

\usepackage{mathptmx}       
\usepackage{helvet}         
\usepackage{courier}        
\usepackage{type1cm}        
\usepackage{makeidx}         
\usepackage{graphicx}        
\usepackage{multicol}        
\usepackage[bottom]{footmisc}

\makeindex             

\begin{document}

\title*{Radiation hydrodynamics simulations of massive star formation using Monte Carlo radiation transfer}
\titlerunning{Massive star formation simulations}
\author{Tim J. Harries, Tom J. Haworth and David Acreman}
\authorrunning{Harries, Haworth \& Acreman}
\institute{Department of Physics and Astronomy, University of Exeter, UK. }
%
%
\maketitle

\abstract*{}

\abstract{We present a radiation hydrodynamics simulation of the formation of a massive star  using a Monte Carlo treatment for the radiation field. We find that  strong, high speed bipolar cavities are driven by the radiation from the protostar, and that accretion occurs stochastically from a circumstellar disc. We have computed spectral energy distributions and images at each timestep, which may in future be used to compare our models with photometric, spectroscopic, and interferometric observations of young massive stellar objects. }

\section{Introduction} 

The deposition of substantial momentum from the radiation field into the accreting material of the protostellar envelope creates a serious obstacle for theories of  massive star formation that adopt spherical accretion e.g. \cite{wolfire_1987}. Extending the models to multiple dimensions and the inclusion of rotation leads to a scenario in which the protostar can accrete via a disc, which subtends a smaller solid angle at the protostellar surface and thus intercepts a smaller fraction of the radiation e.g. \cite{yorke_2002}.

Recent work by Krumholz et al. used adaptive mesh refinement hydrodynamics coupled with a flux limited diffusion (FLD) approximation for the radiative transfer (RT) in order to simulate the formation of a massive binary \cite{krumholz_2009}. They found accretion occurred via a radiatively-driven Rayleigh Taylor instability that allowed material to penetrate the radiation-driven cavities above and below the protostars. Simulations by Kuiper et al. \cite{kuiper_2012} using a hybrid RT method \cite{kuiper_2010} have cast doubt on this scenario -- they claim that the grey FLD approximation leads to a substantial underestimate of the driving force of the radiation field (see also Kuiper's contribution to these proceedings).

Clearly the level of detail with which the radiation transport is followed is an important consideration when simulating radiation feedback. We have therefore developed our own radiation-hydrodynamics code \cite{haworth_2012} which is based on the {\sc torus} Monte Carlo RT code \cite{harries_2000,harries_2004, pinte_2009}. The advantage of this approach is that RT is fully polychromatic and includes both dust and atomic microphysics, and although the method is unavoidably computationally intensive, the  overhead is reduced to acceptable levels by the use of aggressive parallelization.

\section{The Monte Carlo method and its implementation}

The {\sc torus} code is based on the radiative-equilibrium method of \cite{lucy_1999} in which the stellar luminosity is divided up into a large number of photon packets, which propagate through the computational domain undergoing absorptions and scatterings. When a photon packet is absorbed it is immediately re-emitted with a frequency found from random sampling of the emissivity spectrum at that point in the grid. By tracking the paths of the packets through a grid cell one can obtain a Monte Carlo estimate for the energy density in that cell, and hence the mean intensity of the radiation field, which in turn enables the computation of absorption rates, photoionization rates etc. The radiation force on each cell can be computed using similar methods.

The primary drawback in adopting such a detailed treatment of the radiation field is the computational effort involved. Fortunately  the Monte Carlo method, in which the photon packets are essentially independent events, may be straightforwardly and effectively parallelized. The top level of parallelization involves domain decomposing the octree that stores the adaptive mesh. The main bottleneck is the communication overhead when passing photon packets between threads, and we optimize this by passing stacks of photon packet data between the  threads rather than individual packets. A further level of parallelization is achieved by having many sets of identical computational domains, over which the photon packet loop is split, with the results derived from the radiation calculation (radiation momenta, absorption rate estimators etc) collated and returned to all the sets at the end of each iteration. Using these parallelization methods it is possible to reduce the time for the radiation field calculation down to a level comparable with that of the hydrodynamics step.

\section{A test calculation}

We adopted the initial conditions from one of the models presented by \cite{kuiper_2012} in order to perform a demonstration calculation. This model consists of a 100\,M$_\odot$, 0.1\,pc radius cloud with an $r^{-1.5}$ density profile and a small amount ($\Omega = 5 \times 10^{-13}$ s$^{-1}$) of solid body rotation. The effective temperature of the protostar and its luminosity were determined by interpolating  by mass in the protostellar evolution models of Hosokawa \& Omukai \cite{hosokawa_2009}. The photospheric spectrum was composed of an appropriate Kurucz model plus a blackbody accretion spectrum. A $128^3$ regular mesh was used for this test calculation.

\begin{figure}
\sidecaption                                                   
\includegraphics[width=7cm]{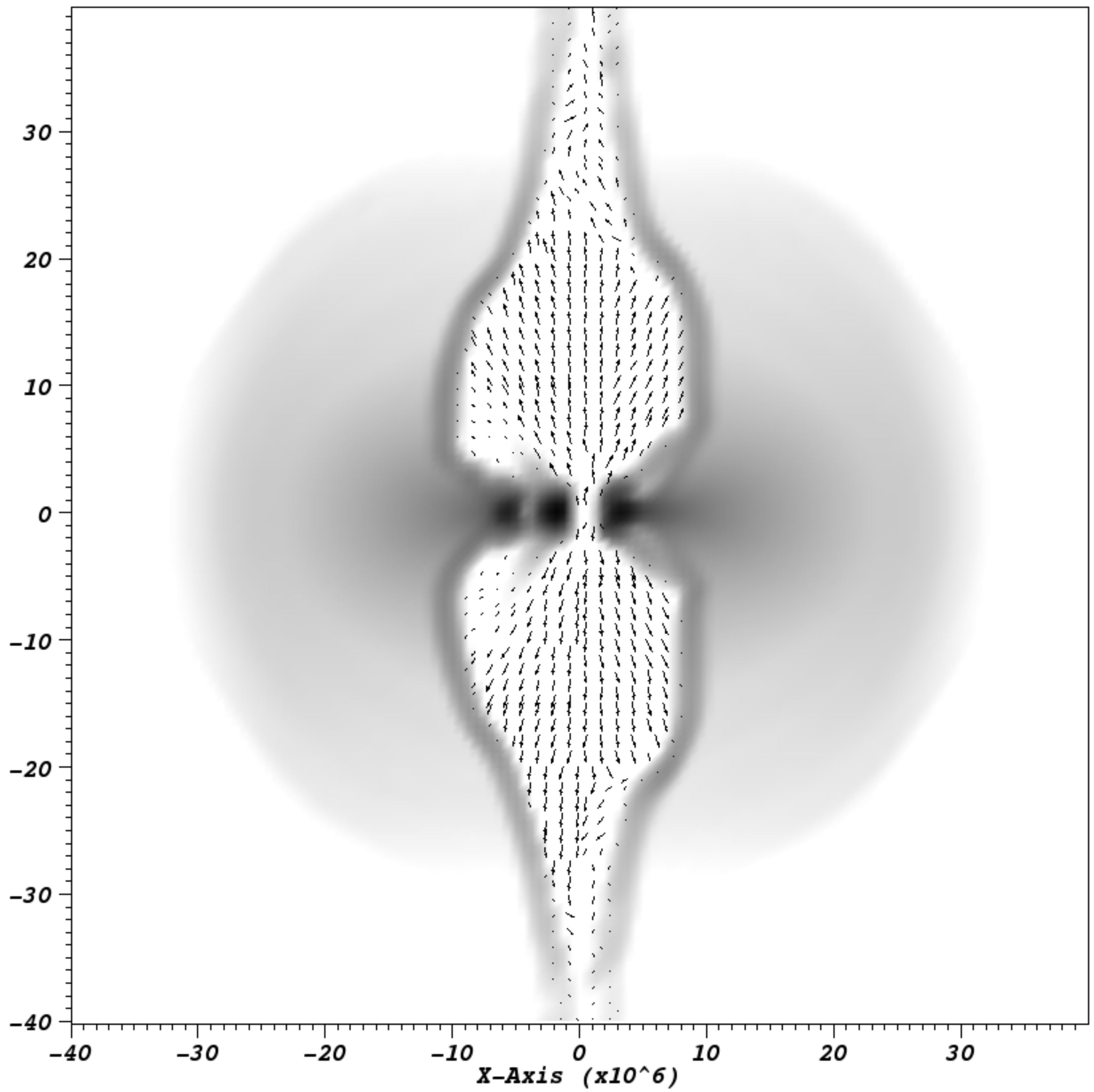}
%
%
\caption{A slice through the rotation axis of the test simulation (spatial units are $10^{10}$ cm). The greyscale shows logarithmic density scaled between 10$^{-19}$ g\,cm$^{-3}$ (white) and 10$^{-15}$ g\,cm$^{-3}$ (black). The arrows denote velocity, with the longest arrows corresponding to speeds of 100 km\,s$^{-1}$.}
\label{harries_fig1}       
\end{figure}

\begin{figure}
\includegraphics[width=\textwidth]{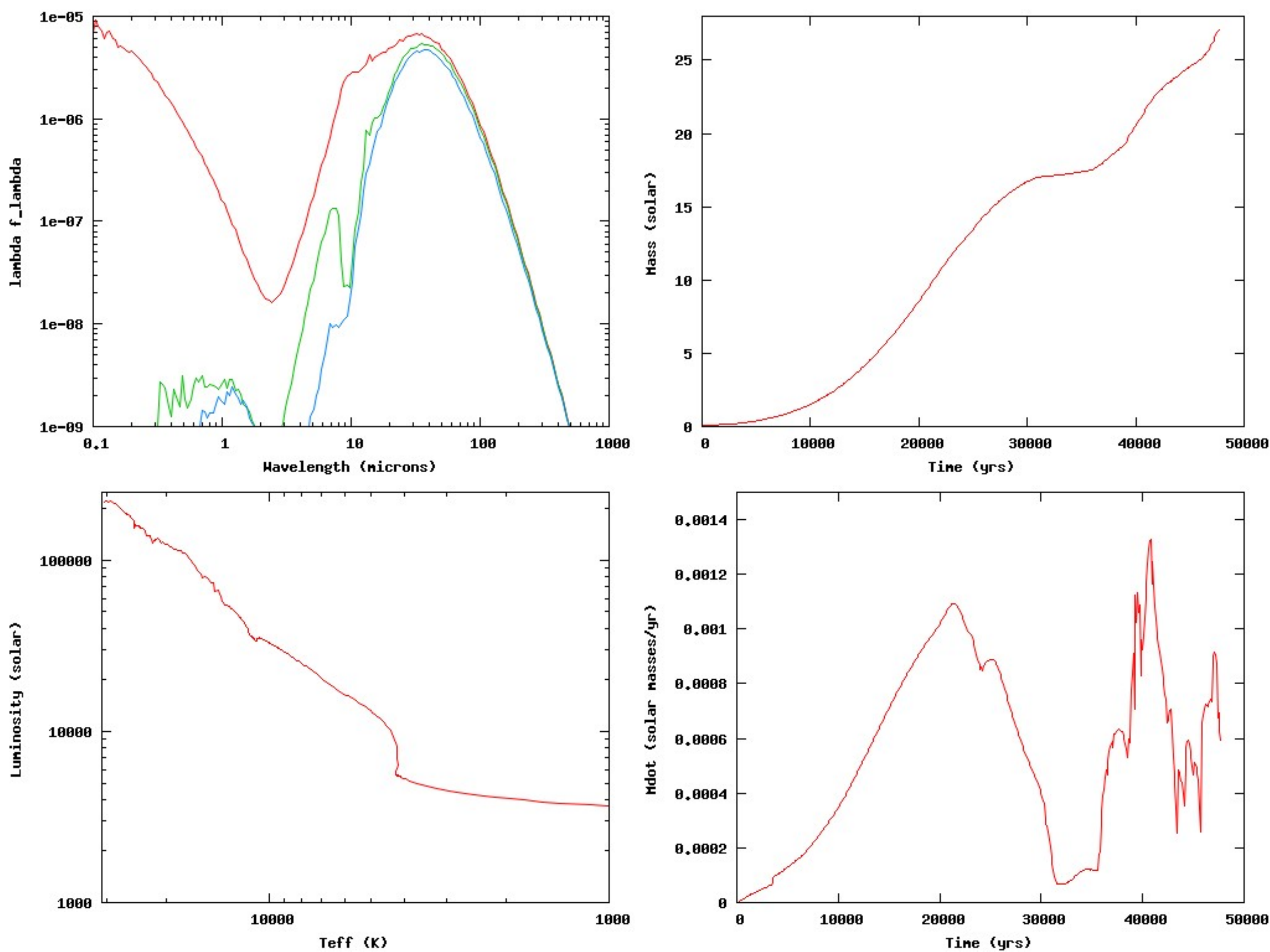}
%
%
\caption{The results of the test calculation. Spectral energy distributions (erg\,s$^{-1}$\,cm$^{-2}$ at a distance of 1\,kpc) at inclinations of 0, 60 and 90 degrees are shown in the top-left figure. The mass of the protostar as a function of time is plotted in the top-right figure. The effective temperature and luminosity of the protostar is shown in the bottom left figure (note the temperature increases to the left). The bottom-right figure shows the accretion rate onto the protostar as a function of time.}
\label{harries_fig2}       
\end{figure}

The final state of the model (after 50\,000 years) is plotted in Fig.~\ref{harries_fig1}. The radiation-driven bipolar cavities are clearly seen, and the outflow has reached the boundary of the computational domain. The development and speed of the cavities is similar to that found by Kuiper et al., but the morphology is different, principally due to the resolution. We have also computed SEDs at each timestep, and plot these in Fig.~\ref{harries_fig2}. Note that 60$^\circ$ and 90$^\circ$ inclination SEDs show typical Class~I  characteristics, and the attenuated protostellar SED is observed only directly along the outflow cavities. The evolution of the accretion onto the protostar can be divided into two distinct phases: At early times the accretion is smooth, and from the envelope, but after $\sim$20\,000 years the luminosity of the protostar overtakes the accretion luminosity (this is dictated by the evolutionary tracks) and the radiation-pressure starts to drive the polar outflow cavities, at which point accretion is occurring stochastically via the protostellar disc. (Note that this two stage accretion evolution is also present in the hybrid RT calculations of Kuiper et al.) At the end of the calculation the protostar has a mass of 28\,M$_\odot$ and an effective temperature of 30\,000 K, at which point ionization and stellar wind feedback are potentially important. Although a small ionized zone is formed close to the protostar towards the end of the calculation (with properties that would identify it as an ultra-compact H\,{\sc ii} region) we note that we have yet to implement a self-consistent method for treating the dust sublimation and advection.

{\noindent \bf Acknowledgements} We thank Takashi Hosokawa for providing digital versions of his evolutionary tracks and Ant Whitworth for a useful discussion whilst floating around in the Med.

\bibliographystyle{spphys}
\bibliography{harries}

\end{document}